\documentclass[sigconf]{acmart}
\AtBeginDocument{%
  } 

\setcopyright{acmlicensed}
\copyrightyear{2026}
\acmYear{2026}
\acmDOI{XXXXXXX.XXXXXXX}
\acmConference{JAWs 2026--ICSE 2026}{APRIL 13--14, 2026}{RIO}
\acmISBN{978-1-4503-XXXX-X/2018/06}

\usepackage{multirow}
\usepackage{array}
\usepackage{booktabs}
\usepackage{makecell}
\usepackage{pifont}

\usepackage{ifthen}
\newboolean{showcomments}
\setboolean{showcomments}{true} 

\newcommand{\todoc}[2]{\ifthenelse{\boolean{showcomments}}{\textcolor{#1}{\textbf{#2}}}{}}

\begin{document}

\title{Weaponizing the Commons: A Taxonomy and Detection Framework of Abuse on GitHub}

\author{Yuli Cheng}
\orcid{0009-0003-7112-449X}
\affiliation{%
  \institution{Xi'an Jiaotong University}
  \country{China}
}
\email{1467959770@stu.xjtu.edu.cn}

\author{Xiaoyu Zhang}
\orcid{0000-0001-7010-6749}
\affiliation{%
  \institution{Nanyang Technological University}
  \country{Singapore}
}
\email{xiaoyu.zhang@ntu.edu.sg}

\author{Jiongchi Yu}
\orcid{0000-0002-2888-4499}
\affiliation{%
  \institution{Singapore Management University}
  \country{Singapore}
}
\email{jcyu.2022@phdcs.smu.edu.sg}

\author{Shiqing Ma}
\orcid{0000-0003-1551-8948}
\affiliation{%
  \institution{University of Massachusetts, Amherst}
  \country{United States}}
\email{shiqingma@umass.edu}

\author{Chao Shen}
\orcid{0000-0002-6959-0569}
\authornote{Chao Shen is the corresponding author.}
\affiliation{%
  \institution{Xi'an Jiaotong University}
  \country{China}}
\email{chaoshen@xjtu.edu.cn}

\author{Yang Liu}
\orcid{0000-0001-7300-9215}
\affiliation{%
  \institution{Nanyang Technological University}
  \country{Singapore}}
\email{yangliu@ntu.edu.sg}

\renewcommand{\shortauthors}{Cheng et al.}

\begin{abstract}
GitHub plays a critical role in modern software supply chains, making its security an important research concern. Existing studies have primarily focused on CI/CD automation, collaboration patterns, and community management, while abuse behaviors on GitHub have received little systematic investigation. In this paper, we systematically review and summarize reported GitHub abuse behaviors and conduct an empirical analysis of publicly available abuse cases, curating a manually labeled dataset of 392 GitHub instances. Based on this investigation, we propose a comprehensive taxonomy that characterizes their diverse symptoms and root causes from a software security perspective. Building on this taxonomy, we develop a unified detection framework capable of identifying all abuse categories across repositories and user accounts. Evaluated on the constructed dataset, the proposed framework achieves high performance across all categories (e.g., F1-score exceeding 89\%). Collectively, this work advances the understanding of GitHub abuse behaviors and lays the groundwork for large-scale, systematic analysis of the GitHub platform to strengthen software supply chain security.
\end{abstract}





\maketitle

\section{Introduction}
Software hosting platforms are foundational to modern software supply chains, supporting code contribution, version control, dependency management, and CI/CD pipelines~\cite{wenEmpiricalStudySecurity2020, valenzuela-toledoHiddenCostsAutomation2024, liTechnicalDebtManagement2022}. Among them, GitHub is the most widely used platform globally, with extensive user activity and development~\cite{limaCodingTogetherScale2014, saroarGitHubMarketplacePractitioners2022, alshomali2017github}. Investigating security issues on GitHub is therefore of paramount importance for ensuring the integrity and security of the software supply chain.

Existing research on GitHub security has primarily focused on development workflows and CI/CD automation~\cite{valenzuela-toledoHiddenCostsAutomation2024, panAmbushAllSides2024}, developer behavior and collaboration patterns~\cite{alrashedyHowSoftwareEngineering2024, batounEmpiricalStudyGitHub2023}, software quality and engineering practices~\cite{wangAccuracyEmpiricalStudy2024, liTechnicalDebtManagement2022}, and related aspects.
However, abuse behaviors on GitHub, such as activities that intentionally distort platform signals, disrupt normal development processes, or deceive users, have received limited systematic research attention.
Existing studies have mostly examined isolated issues such as fake stars~\cite{heSixMillionSuspected2025}, without offering a comprehensive understanding of the diverse forms, symptoms, and impacts of abuse behaviors on GitHub.

Unlike software vulnerabilities, abuse behaviors exploit the platform's open collaboration and low-barrier-to-entry features and can pose multi-layered, chain-like risks to the software ecosystem.
These abuse behaviors can erode user trust in projects and interactions and disrupt normal development workflows (e.g., interfering with issue and code management), leading to high governance costs and low collaboration quality~\cite{mcmahonGitHubBlackMarket, majumdarMillionsFakeRepositories2024, gonzalezAnomaliciousAutomatedDetection2021}.
Even worse, when attackers deliberately leverage platform mechanisms to spread malicious artifacts or deliver harmful content, the resulting impact can propagate from a single repository to a broader software supply chain, ultimately threatening the security of users and production~\cite{NewTechniqueDetected, newsHackersAreAbusing2024, thomas2021sok, heSixMillionSuspected2025}.
Therefore, there is an urgent need to conduct a comprehensive investigation and understand the breadth, characteristics, and symptoms of abuse behaviors on GitHub.


To fill this gap, this work systematically investigates eight types of GitHub abuse behaviors by manually reviewing and summarizing relevant literature and reports. Based on this investigation, we propose a comprehensive taxonomy, categorizing the abuse behaviors into four high-level classes according to their impact, and provide detailed descriptions and characteristic symptoms for each type. Building on this taxonomy, we further design a unified detection framework capable of identifying specific abuse behaviors. The proposed framework is evaluated on a labeled dataset consisting of 392 instances, achieving F1-score exceeding 89\% across multiple abuse categories. The main contributions are summarized as follows:
\begin{itemize}
    \item We propose a comprehensive taxonomy of eight types of abuse behaviors on GitHub, along with detailed descriptions of their observable symptoms and root causes.
    \item Building on this taxonomy, we design a unified detection framework. Experimental results on a labeled dataset of 392 instances demonstrate its effectiveness. 
    \item We publicly release the source code of the proposed framework\footnote{\url{https://github.com/Rasnd-yu/GitHub-Detector}}, serving as a foundation for future studies on abuse behaviors and detection techniques on GitHub.
\end{itemize}

\section{Background and Related Works}
\textbf{Open-Source Community Security.} With the increasing reliance on open-source software, security issues in open-source communities have attracted sustained research attention. As the world's largest open-source hosting platform, GitHub has been studied from multiple security perspectives. Prior work has explored the detection of anomalous or malicious commits using repository metadata~\cite{gonzalezAnomaliciousAutomatedDetection2021}, identified large-scale fake starring behaviors~\cite{heSixMillionSuspected2025}, and proposed automated approaches for detecting spam or low-quality content in issue discussions~\cite{firake2025spamfilter}.

Beyond GitHub-specific studies, a broader body of research examines security practices, risks, and governance challenges in OSS ecosystems~\cite{wenEmpiricalStudySecurity2020, schliengerAnalyzingInformationSecurity2003}. Software supply-chain security, in particular, has become a central research focus, with extensive work analyzing attack vectors and corresponding defenses~\cite{ohmBackstabbersKnifeCollection2020, williamsResearchDirectionsSoftware2025, okaforSoKAnalysisSoftware2022}. Despite these efforts, systematic investigation of socio-technical abuse behaviors within GitHub remains limited. In particular, there is still no unified taxonomy or detection framework that captures GitHub abuse behaviors, an important gap this work aims to address.

\noindent
\textbf{Online Misbehaviors.} Online misbehavior has been extensively studied across diverse online platforms, including social networks, financial communities, and collaborative systems. In social networking contexts, prior work has proposed comprehensive taxonomies of harmful behaviors such as hate, harassment, and fake accounts, and analyzed their associated user groups and behavioral patterns~\cite{thomas2021sok, fireFriendFoeFake2014}. A long line of research on Sybil attacks further investigates the structural properties of inauthentic identities and develops effective detection mechanisms~\cite{wangSocialTuringTests2012, alvisiSoKEvolutionSybil2013}. As a result, a variety of mature techniques now exist for detecting spam, fake users, and inauthentic activities in online social networks~\cite{yuanDetectingFakeAccounts2019, uddinSpammerDetectionFake2019}.

More recent studies emphasize that online misbehavior is often intertwined with platform-specific trust and reputation signals. For instance, evidence shows that profile signals and verification mechanisms can be manipulated or misinterpreted across platforms~\cite{cuevas2025profile}. Similarly, research on domain-specific communities, such as investment platforms, demonstrates that misbehavior manifests differently depending on the underlying social and reputational structures, requiring tailored analysis and mitigation strategies~\cite{tsuchiyaMisbehaviorAccountSuspension2023}. Collectively, these studies suggest that online misbehavior is not solely content-driven, but is deeply embedded in platform-dependent socio-technical mechanisms.

\section{A Taxonomy of GitHub Abuse Behaviors}
\subsection{Overview of the Taxonomy}
GitHub provides core functionalities including search, automation, project development, and collaborative interactions. Analyzing how abuse behaviors affect these functionalities, we construct a taxonomy with four high-level categories and eight subcategories (Table~\ref{tab_taxonomy}). Each category is accompanied by detailed descriptions of characteristic symptoms, enabling a systematic understanding of how different abuse behaviors affect GitHub's key functionalities.

\renewcommand{\arraystretch}{1.2}
\begin{table*}[t]
  \caption{Taxonomy of Abuse Behaviors on GitHub.}
  \label{tab_taxonomy}
  \begin{tabular}{ccm{10.2cm}}
    \toprule
    Category & Subcategory & \makecell[c]{Symptom} \\
    \midrule

    \multirow{6}{*}{\centering Attention Hijacking}
    & Fake Stars
    & \(
    \displaystyle
    ST(u,r) = 1
    \;\wedge\;
    |\mathcal{SR}_u| \le x_1
    \;\wedge\;
    AT(u,[t_s,\,t_s+\Delta t]) \le \epsilon
    \) \\
    \addlinespace[0.5ex]
    \cline{2-3}
    \addlinespace[0.5ex]
    
    & \multirow{1.2}{*}{Automatic Updates}
    & \(
    \displaystyle
    CM(r,[t_c, t_c+\Delta t]) \ge x_2
    \;\wedge\;
    \frac{LOC_m(r,[t_c, t_c+\Delta t])}{CM(r,[t_c, t_c+\Delta t])} \le y
    \) \\
    \addlinespace[0.5ex]
    \cline{2-3}
    \addlinespace[0.5ex]
    
    & Keyword Stuffing 
    & \(
    \displaystyle
    \Big|\Big\{
    k \in \mathcal{K}_r
    \;\big|\;
    Rel(k,\mathrm{RD}_r) < \theta_k
    \Big\}\Big| \ge x_3
    \) \\
    \addlinespace[0.5ex]
    \cline{2-3}
    \addlinespace[0.5ex]
    
    & Typo Squatting
    & \(
    \displaystyle
    Sim(n_i, n_j) \ge \theta_{t1}
    \;\wedge\;
    Sim(\mathrm{RD}_i, \mathrm{RD}_j) \ge \theta_{t2} 
    \;\wedge\;
    RT\Big(P(r_i),P(r_j)\Big) \ge \phi_{p1} 
    \) \\
    \midrule

    \multirow{1}{*}{\centering Authority Fraud}
    & Spoofed Contributor 
    & \(
    \displaystyle
    \Big|\Big\{c \in \mathcal{C}_r \mid Ath(c)=u\Big\}\Big| \le x_4
    \;\wedge\;
    P(r) \le \phi_{p2}
    \;\wedge\;
    P(u) \ge \phi_{p3}
    \) \\
    \midrule

    \multirow{1}{*}{\centering Spam}
    & Issue Spam 
    & \(
    \displaystyle
    i \in \mathcal{I}_r
    \;\wedge\;
    \Big( \big(HL(i) = 1\big) \vee \big(HC(i) = 1\big) \Big)
    \;\wedge\;
    SP(i) = 1
    \) \\
    \midrule
    
    \multirow{2.5}{*}{\centering Reputation Manipulation}
    & Reputation Farming 
    & \(
    \displaystyle
    a \in \mathcal{I}_r \cup \mathcal{PR}_r
    \;\wedge\;
    AT(u, [t_r(a)+\delta_t, t_r(a)+\Delta t]) \ge 1
    \) \\
    \addlinespace[0.5ex]
    \cline{2-3}
    \addlinespace[0.5ex]
    
    & Fake Stats 
    & \(
    \displaystyle
    \Big( \exists r \in \mathcal{R}_{\text{others}} : url_r \in \mathcal{L}_u \Big)
    \;\vee\;
    \Big( \bigl| CS(u) - \sum\nolimits _r S(r) \bigr| \ge x_5, r \in \mathcal{R}_u \Big)
    \) \\
    \bottomrule
  \end{tabular}
  \scriptsize
  \begin{minipage}[t]{0.27\textwidth}
  \begin{tabular}{rm{3.9cm}}
    1 & $ST/HL/HC/SP$, boolean function \\
    2 & $ST(u,r)$, user $u$ starred on repositories $r$ \\
    3 & $\mathcal{SR}_u$, set of repositories starred by user $u$ \\
    4 & $AT(u,T)$, user $u$'s activity count over $T$ \\
    5 & $t_s$, timestamp of star activity \\
    6 & $x_1/x_2/x_3/x_4/x_5$, count threshold \\
    7 & $\epsilon$, inactivity threshold \\
    8 & $CM(r,T)$, repository $r$'s total commit over $T$ \\
    9 & $LOC_m(r,T)$, total modified lines of code in repository $r$ over $T$ \\
    10 & $t_c$, timestamp of commit activity \\
  \end{tabular}
  \end{minipage}
  \hfill  
  \begin{minipage}[t]{0.24\textwidth}
  \begin{tabular}{rm{3.4cm}}
    11 & $y$, average quantity threshold \\
    12 & $\mathcal{K}_r$, set of keywords in the repository $r$ \\
    13 & $Rel(t,T)$ the relevance score between a short text $t$ and a long text $T$ \\
    14 & $\mathrm{RD}_r$, the README content of GitHub repository $r$ \\
    15 & $\theta_k$, relevance threshold \\
    15 & $\theta_{t1}/\theta_{t2}$, similarity thresholds \\
    16 & $Sim(\alpha,\beta)$, text $\alpha$-$\beta$ similarity score \\
    17 & $n_i/n_j$, name of GitHub repository \\
    18 & $\phi_{p1}/\phi_{p2}/\phi_{p3}$, popularity thresholds \\
  \end{tabular}
  \end{minipage}
  \hfill
  \begin{minipage}[t]{0.23\textwidth}
  \begin{tabular}{rm{3.2cm}}
    19 & $RT(a,b)$, popularity ratio defined as $\max(a,b)/\min(a,b)$ \\
    20 & $\mathcal{C}_r$, set of commits of repository $r$ \\
    21 & $P(r)/P(u)$, normalized popularity metric (e.g., stars, forks) \\
    22 & $Ath(c)$, author of commit $c$ \\
    23 & $\mathcal{I}_r$, set of issues in repository $r$ \\
    24 & $HL(i)/HC(i)$, issue $i$ contains one or more links/commands \\
    25 & $SP(i)$, issue $i$ contains spam or phishing-like content \\
  \end{tabular}
  \end{minipage}
  \hfill
  \begin{minipage}[t]{0.23\textwidth}
  \begin{tabular}{rm{3cm}}
    26 & $\mathcal{R}_u$, set of repositories of user $u$ \\
    27 & $\mathcal{PR}_r$, set of pull requests associated with repository $r$ \\
    28 & $t_r(a)$, timestamp when $a$ (issue or PR) is closed or merged \\
    29 & $\delta_t$, delay threshold after $a$ (issue or PR) closure/merge \\
    30 & $\mathcal{L}_u$, set of README stat URLs used by user $u$ on profile \\
    31 & $CS(u)$, user $u$'s claimed star count \\
    32 & $S(r)$, star count of repository $r$ \\
  \end{tabular}
  \end{minipage}
\end{table*}

\subsection{Attention Hijacking}
\emph{Attention Hijacking} refers to the abuse of GitHub's repository discovery and search mechanisms. When searching for repositories, users commonly rely on signals such as popularity indicators (e.g., stars and forks), update recency, and keyword-based relevance \cite{borgesPopularityGitHubApplications2017}. Abusers exploit these signals to artificially elevate repository rankings and visibility, thereby misleading users or promoting low-quality or malicious projects. Based on the exploited signals, we further divide \emph{Attention Hijacking} into four representative subcategories: \emph{Fake Stars}, \emph{Automatic Updates}, \emph{Keyword Stuffing}, and \emph{Typo Squatting}, as summarized in~\autoref{tab_taxonomy}.

\noindent
\(\bullet\)
\textbf{Subcategory 1 (Fake Stars).}
Abusers purchase or trade fake stars to rapidly inflate a repository's apparent popularity and reputation~\cite{mcmahonGitHubBlackMarket, universityFraudstersUseFakedateFormat}. These artificially boosted repositories are more likely to attract developers or investors, and in some cases are used to lure unsuspecting users into scams, credential theft, cryptocurrency fraud, or the distribution of malicious software.

\noindent
\(\bullet\)
\textbf{Subcategory 2 (Automatic Updates).}
Abusers leverage GitHub Actions to automatically push trivial updates at a high frequency, often by repeatedly modifying a file which is usually called ``log''~\cite{newsBewareGitHubsFake, NewTechniqueDetected}. This activity creates the illusion of active maintenance and disproportionately increases repository visibility, particularly when users sort search results by recent updates.

\noindent
\(\bullet\)
\textbf{Subcategory 3 (Keyword Stuffing).}
Repositories are populated with a large number of popular or trending keywords to manipulate search rankings~\cite{newsBewareGitHubsFake, newsHackersAreAbusing2024}. These keywords are often unrelated to the repository name or README content, misleading users about the repository's actual purpose or functionality.

\noindent
\(\bullet\)
\textbf{Subcategory 4 (Typo Squatting).}
\emph{Typo Squatting} involves creating repositories with names that closely resemble those of popular projects, exploiting common typographical errors or minor naming variations~\cite{yakobiWatchTypoOur2024, majumdarMillionsFakeRepositories2024}. Users may mistakenly assume these malicious repositories as legitimate or widely used projects.

\subsection{Authority Fraud}
\emph{Authority Fraud} targets the perceived credibility of GitHub repositories by exploiting the reputational capital of well-known developers. Rather than manipulating popularity or visibility signals, attackers seek to inflate a project's apparent trustworthiness, misleading users about its legitimacy or quality. In our study, the predominant form of \emph{Authority Fraud} on GitHub is \emph{Spoofed Contributor}.

\noindent
\(\bullet\)
\textbf{Subcategory 5 (Spoofed Contributor).}
GitHub allows repository owners to attribute co-authorship via email addresses in commit messages~\cite{CreatingCommitMultiple}. Malicious actors exploit this mechanism by impersonating well-known developers as co-authors to enhance the perceived legitimacy and reputation~\cite{FakeGitHubCommits}. As GitHub does not notify users when they are listed as co-authors, the impersonation may remain unnoticed, increasing the stealthiness of this abuse.

\subsection{Spam}
The \emph{Spam} category encompasses abuse behaviors that introduce unwanted or harmful content into GitHub interaction channels, such as irrelevant or malicious posts, bogus pull requests, and automated notifications. Such spam is increasingly generated by automated agents and bot-like accounts, creating noise that developers must filter as part of regular project maintenance \cite{henselSurveyAutomatedAgents}. In this work, we focus on spam appearing in repository issue trackers, as it directly affects communication between contributors and can degrade the quality of project interaction and developer experience.

\noindent
\(\bullet\)
\textbf{Subcategory 6 (Issue Spam).}
The content of issues is unrelated to the target repository and instead contains phishing messages, online scams, or links to malicious software. In some cases, abusers automate issue creation using GitHub Actions to distribute deceptive messages (e.g., fake “IMPORTANT” notifications), facilitating large-scale phishing or fraud campaigns~\cite{CleverGitHubScanner, ThisWindowsPowerShell2024}.

\begin{figure*}[t]
  \centering
  \includegraphics[width=\textwidth]{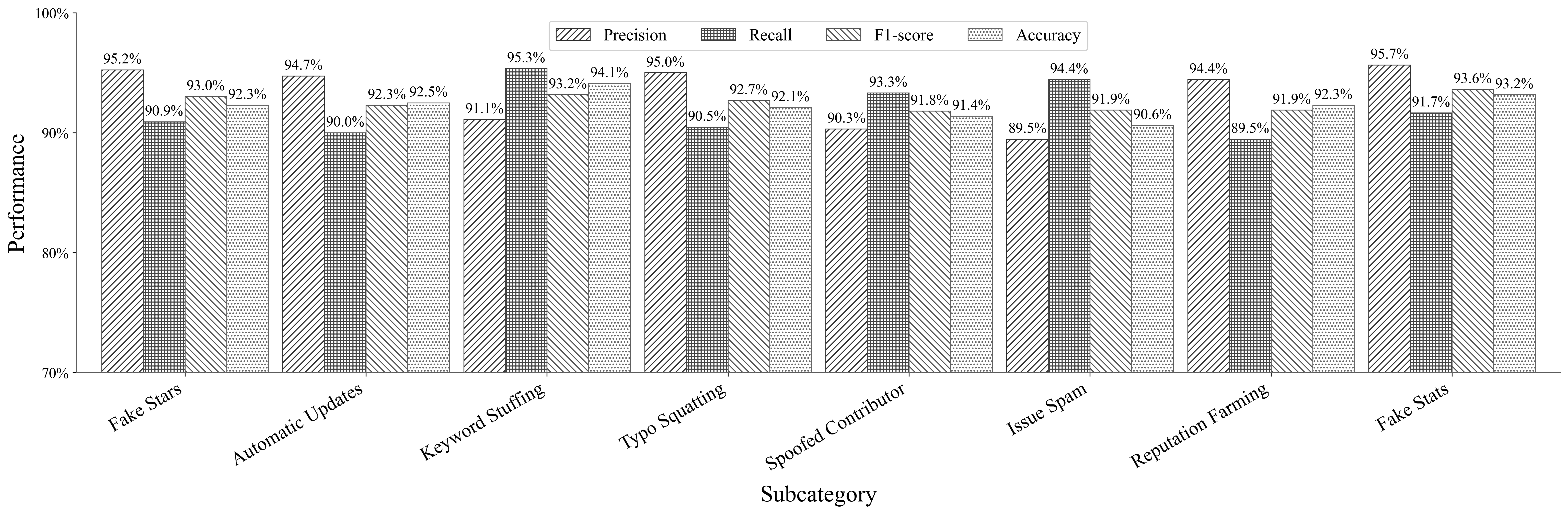}
  \caption{The performance of the comprehensive detection framework on small-scale artificial datasets}
  \Description{Comprehensive performance comparison}
  \label{framework_results}
\end{figure*}

\subsection{Reputation Manipulation}
On GitHub, each user maintains a public profile representing personal identity and reputation. Profiles expose signals such as contribution history, activity records, and achievements, which influence perceived expertise, credibility, and trustworthiness~\cite{READMEBadgesAre2018}. However, these signals are not always verifiable, and some users manipulate or fabricate them to artificially enhance influence or visibility. We categorize such behaviors as \emph{Reputation Manipulation}, comprising two subcategories: \emph{Reputation Farming} and \emph{Fake Stats}.

\noindent
\(\bullet\)
\textbf{Subcategory 7 (Reputation Farming).}
\emph{Reputation Farming} inflates apparent user activity by performing low-effort interactions, such as approving or commenting on pull requests and issues that have already been resolved or closed~\cite{OpenSSFWarnsReputation, ashwinReputationFarmingOSS2024}. These actions contribute little substantive value, yet are prominently recorded in profile activity timelines, creating a misleading impression of sustained and meaningful participation. 

\noindent
\(\bullet\)
\textbf{Subcategory 8 (Fake Stats).}
Profile-level manipulation can be used to spoof personal credibility and mislead users into trusting a contributor~\cite{koldobsky5WaysAttackers2023}. Such manipulation includes falsely claiming membership in well-known organizations, displaying fabricated achievement badges, or embedding third-party statistic widgets (e.g., \texttt{github-readme-stats}) that reference another user's account.

\section{Framework \& Experiment}
Based on the taxonomy, we design a unified detection framework that integrates multiple detection strategies tailored to different abuse symptoms. Its effectiveness is evaluated using a small-scale labeled dataset containing 392 instances spanning all categories, and the results are summarized in~\autoref{framework_results}.

\subsection{Setup}
\textbf{Framework Implementation.}
For certain abuse categories, we directly adopt established detection approaches. Specifically, \emph{Fake Stars} are detected using \emph{StarScout}~\cite{heSixMillionSuspected2025}, which identifies anomalous star-giving behavior. For \emph{Issue Spam}, we apply a MLP classifier with TF-IDF vectorizer~\cite{firake2025spamfilter}, designed for issue content classification.

For the remaining categories, we design tailored detection strategies based on the symptoms summarized in our taxonomy and insights from related studies. For \emph{Keyword Stuffing}, we leverage the well-established BM25 ranking algorithm to measure the relevance between repository metadata and injected keywords, enabling the identification of anomalous keyword usage. Similarly, \emph{Typo Squatting} is detected using mature text similarity metrics to identify pairs of repositories with highly similar names but significantly different popularity levels, indicating potential impersonation.

Detection of \emph{Automatic Updates} exploits characteristic temporal patterns. We first retrieve repositories with frequent recent updates via the GitHub API to narrow the candidate set, and then analyze commit frequency and code change quality to distinguish abusive update behavior from legitimate maintenance activity. The remaining categories, \emph{Spoofed Contributor}, \emph{Reputation Farming}, and \emph{Fake Stats}, require more fine-grained and resource-intensive analysis. For example, for \emph{Spoofed Contributor}, we analyze the activity patterns of reputable contributors to determine whether their identities have been exploited by other actors. In the case of \emph{Reputation Farming}, we conduct detailed activity-level analyses of target contributors, focusing on interactions with already closed or resolved pull requests and issues. Detection of \emph{Fake Stats} requires examining individual claims and statements presented on user profile pages and verifying their consistency with ground-truth data. 

\textbf{Dataset Structure.}
As part of our systematic investigation of abuse behaviors on GitHub, we constructed a labeled dataset of 392 abuse instances covering the period since 2020, composed of 310 GitHub repositories and 82 user accounts. The dataset encompasses repositories across industrial software, e-commerce platforms, and both front-end and back-end projects, as well as users exhibiting abuse behaviors in commits, pull requests, issues, and other collaborative interactions. Each instance is manually annotated with its abuse subcategory, and the dataset is balanced across positive and negative samples. Sources include prior academic studies, security reports, and gray literature.

\subsection{Result Analysis}
As shown in Figure~\ref{framework_results}, the framework consistently achieves strong performance across all abuse categories, with all metrics exceeding 89\%. Notably, \emph{Keyword Stuffing} and \emph{Fake Stats} attain the highest F1-scores (93.2\% and 93.6\%), largely due to the maturity of the BM25 algorithm and the structured nature of these symptoms. In contrast, \emph{Issue Spam} and \emph{Reputation Farming} show slightly lower performance with larger Precision-Recall gaps, likely due to their diverse and irregular symptoms, while their lowest metrics still reach 89.5\%, keeping the results within an acceptable range. Overall, the framework demonstrates robust performance, suggesting that the task's structured and regular nature allows a rule-based approach to achieve strong results, obviating the necessity for more sophisticated models.

\section{Conclusion and Future Work}
In this paper, we present a systematic investigation of abuse behaviors on GitHub by reviewing existing reports and prior research into a comprehensive taxonomy. Building on this taxonomy, we design a unified detection framework that identifies a diverse range of abuse behaviors. Experimental results on a labeled dataset demonstrate the effectiveness of the proposed framework.

This work represents an initial step toward systematically characterizing abuse behaviors on GitHub. Looking ahead, we identify two main directions for future research.
\ding{182} {\it Large-scale analysis.} The current manually curated dataset is sufficient for validation but limited in scale. We plan to develop a scalable and automated scanning system to extend detection across broader portions of the GitHub ecosystem, enabling continuous data collection, longitudinal studies, and more accurate estimation of abuse prevalence and evolution.
\ding{183} {\it Deeper empirical insights.} Large-scale scanning will facilitate fine-grained analyses, such as examining correlations and co-occurrence patterns among abuse categories, identifying temporal trends, and uncovering rare or emerging forms of abuse. These efforts may further refine the proposed taxonomy and contribute to a deeper and more comprehensive understanding of socio-technical abuse dynamics in open-source ecosystems.

\newpage


\bibliographystyle{ACM-Reference-Format}
\bibliography{small_reference}

\end{document}